\documentclass[12pt]{iopart}

\usepackage{iopams}  

\usepackage{graphicx}
\usepackage{color}

\begin{document}

\def\tr{\text{tr}}
\def\Tr{\mbox{Tr}}

\newtheorem{theo}{Theorem}

\title{Quantum Fluctuation Relations for Ensembles of Wave Functions}
\author{Michele Campisi}
\address{Institute of Physics, University of Augsburg,
  Universit\"atsstr. 1, D-86135 Augsburg, Germany}
\date{\today }

\begin{abstract}
New quantum fluctuation relations are presented. In contrast with the the standard approach, where
the initial state of the driven system is described by the (micro)canonical density matrix, here we assume
that it is described by a (micro)canonical distribution of wave functions, as originally proposed by Schr\"odinger.
While the standard fluctuation relations are based on von Neumann measurement postulate,
these new fluctuation relations do not involve any quantum collapse,
but involve instead a notion of work as the change in expectation of the Hamiltonian. 
\end{abstract}

\maketitle

\section{Introduction}In the last two decades the field of non-equilibrium thermodynamics has 
undergone a tremendous advancement due to the discovery of exact non-equilibrium
relations (named fluctuation relations) which characterize non-equilibrium processes well beyond the regime of linear response,
and provide a deep insight into statistical nature and microscopic origin of the second law of thermodynamics.
The most prominent example of such exact relations is the Jarzynski equality \cite{Jarzynski97PRL78} which allows
for obtaining the free energy landscape of small systems, like a single DNA molecule, from very many 
measurements of work done on the system as it is driven out of equilibrium, e.g., by stretching the molecule 
\cite{Jarzynski11ARCMP2}. A related result, known as Crooks work fluctuation theorem \cite{Crooks99PRE60}, 
relates the free energy to the  probability of performing work $W$ during the process and the probability
of performing work $-W$ during the time-reversed process. 
These results, which have been first obtained within the 
framework of classical mechanics were later derived also within the quantum mechanical framework
\cite{Kurchan00arXiv,Tasaki00arXiv,Esposito06PRE73,Talkner07JPA40}. 

The crucial ingredient needed for obtaining the fluctuation relations
in the quantum case is the so called {\it two measurements scheme} \cite{Esposito09RMP81,Campisi11RMP83}.
In this scheme the system energy is measured at the beginning and end of the driving protocol and the work
is defined as the difference of the outcomes of these measurements:
\begin{equation}
W= E_m^\tau -E_n^0
\label{eq:w-two-mess}
\end{equation}
where $E_k^t$ denotes an eigenvalue of the (time-dependent) Hamilton operator $\hat H(\lambda_t)$ at time $t$.
As usual, here it is assumed the Hamilton operator changes in time due to the time dependence of an external
parameter $\lambda_t$.
This scheme relies on the von Neumann measurement postulate according to which the measurement 
process induces the collapse of the wave function on one of the eigenstates of the measured observable, i.e,
$\hat H(\lambda_0)$ and $\hat H(\lambda_\tau)$ in the present case. Notably, experimental verification and application 
of the quantum fluctuation relations based on the two-measurement scheme have not been accomplished yet,
while alternative strategies aimed at avoiding the two projective measurements have been proposed.
Two prominent examples propose to replace them with many weak measurements during the driving protocol
\cite{Campisi10PRL105}, or with state tomography of  one or two qubit ancillae appropriately coupled to the
driven system \cite{Dorner13PRL110,Mazzola13PRL110,Campisi13barXiv,Bathalao13arXiv}

With this work we establish {\it new} quantum fluctuation relations, which look exactly like the 
standard quantum fluctuation relations but substantially differ from them due to a different underlying definition of quantum work, 
and a different ensemble specifying the initial condition.
As in the standard case \cite{Esposito09RMP81,Campisi11RMP83} we assume an initial statistical ensemble, but at variance with the 
ordinary quantum statistical mechanics, we assume that the statistical ensemble is described by a 
distribution of wave functions as originally suggested by Schr\"odinger \cite{SchrodingerBook},
later pursued by Khinchin \cite{Khinchin51Book}
and recently advocated by an increasing number of authors 
\cite{Brody98JMP98,JonaLasinio00InBook,Bender05JPA38,Goldstein06JSP125,Goldstein06PRL96,Popescu06NATPHYS2,Naudts06JSM06,Brody07PRSA463,Reimann08PRL101,Reimann07PRL99,Fine09PRE80,Ji11PRL107,Alonso11JPA39}.
We will establish fluctuation relations for the microcanonical  
\cite{Bender05JPA38,Naudts06JSM06,Reimann07PRL99,Reimann08PRL101,Fine09PRE80,Ji11PRL107}
and canonical \cite{Brody98JMP98,Jona-Lasinio06AIPCP844} wave function ensembles.
Most remarkably these new fluctuation relations naturally involve a notion of work as the change in the {\it expectation}
of the Hamiltonian operator
\begin{equation}
w= \langle \psi(\tau) | \hat H(\lambda_\tau) | \psi (\tau)\rangle -  \langle \psi(0) |\hat H(\lambda_0) | \psi(0)\rangle \, .
\label{eq:w-nomeas}
\end{equation}
Accordingly they do not involve von Neumann measurement postulate. In Eq. (\ref{eq:w-nomeas}) $|\psi (0)\rangle$ is a wave function randomly chosen from the distribution,
and $ |\psi (\tau)\rangle$ is its time evolution. Just as with the classical fluctuation theorems, the stochastic 
nature of $w$ comes from the fact that the initial state $|\psi (0)\rangle$ is randomly drawn from a distribution, while
its evolution is deterministic.

So, the interpretation framework that is adopted here is that experimentally observed quantities correspond to
their quantum mechanical {\it expectation}, an approach that is at least as common in the scientific literature and effective as 
that involving wave function collapses. To give one example, Kubo's linear response theory \cite{Kubo57aJPSJ12}, is a theory of quantum 
expectations which mentions no collapses. This same philosophy has been advocated by G. Jona-Lasinio and C. Presilla
\cite{Jona-Lasinio06AIPCP844}, who pointed out that the wave function ensembles could be good candidates for
the study of mesoscopic systems, where robust coherence phenomena are involved.

\section{Wave function ensembles}
We consider a quantum system with a finite dimensional Hilbert space of dimension $N$.
Each wave function $|\psi \rangle$ can be represented by an $N$ dimensional complex vector 
$\mathbf c$, and the system Hamilton operator $\hat H(\lambda)$ can be represented by a $N \times N$ Hermitean matrix 
$\mathbf H(\lambda)$. Following Ref. \cite{Strocchi66RMP38} we introduce the suggestive notation 
\begin{eqnarray}
\mathbf c = \mathbf x + i\mathbf p \\
h(\mathbf x, \mathbf p; \lambda_t) = ( \mathbf x - i\mathbf p)^T \mathbf H(\lambda_t)  ( \mathbf x + i\mathbf p)
\end{eqnarray}
where $\mathbf x$ and $\mathbf p$ are the real and imaginary parts of $\mathbf c$,
 $h(\mathbf x, \mathbf p, \lambda_t)$ denotes the expectation of the Hamilton operator on the state 
$\mathbf x + i\mathbf p$, $\mathbf z^T$ denotes transpose of $\mathbf z$, and matrix multiplication is implied.
We stress that $\mathbf x$ and $\mathbf p$ should not be confused with positions and momenta.

Below we shall consider statistical ensembles $\rho(\mathbf x, \mathbf p)$ defined on the wave function ``phase space'' $(\mathbf x, \mathbf p)$. Given an observable $\hat B$, with matrix representation $\mathbf B$, its ensemble average is its wave function expectation 
$b(\mathbf x, \mathbf p) = ( \mathbf x - i\mathbf p)^T \mathbf B  ( \mathbf x + i\mathbf p)$, averaged over the wave function distribution, namely
\begin{equation}
\langle \hat B \rangle = \int d\mathbf x d\mathbf p\,  \rho(\mathbf x, \mathbf p) b(\mathbf x, \mathbf p) \,.
\end{equation}

\subsection{Microcanonical wave function ensemble}
In the microcanonical wave function ensemble \cite{Bender05JPA38,Naudts06JSM06,Reimann07PRL99,Reimann08PRL101,Fine09PRE80,Ji11PRL107}
all wave functions $(\mathbf x, \mathbf p)$
with a given expectation of energy $E=h(\mathbf x, \mathbf p; \lambda)$ have same 
probability, whereas all other wave functions have probability zero. For a fixed $\lambda$
it reads:
\begin{eqnarray}
\rho_{\mu}(\mathbf x, \mathbf p; E, \lambda)= \frac{\delta(E-h(\mathbf x, \mathbf p; \lambda))
 \delta(1- |\mathbf x + i\mathbf p|^2 )}{\Omega(E,\lambda)}
\end{eqnarray}
where $\delta$ denotes Dirac's delta function, and
\begin{eqnarray}
\Omega(E,\lambda) = \int d\mathbf x d\mathbf p\,  \delta(E-h(\mathbf x, \mathbf p; \lambda))
 \delta(1- |\mathbf x + i\mathbf p|^2 )
 \label{eq:microcan}
\end{eqnarray}
is the density of states. Note the formal similarity with the classical microcanonical ensemble.
The main the difference is the presence of the extra factor $\delta(1- |\mathbf x + i\mathbf p|^2 )$
which restricts the integration to the ``physical Hilbert space'', namely the subspace of normalized wave functions,
also known as the projective Hilbert space.
Note that at variance with the textbook quantum microcanonical ensemble \cite{Huang87Book,Kubo65book},
in which only those eigenstates of the Hamiltonian in a narrow interval around the energy $E$ contribute,
here all eigenstates participate to the ensemble.\footnote{To see this, consider for example a spin-$1$ particle in a (possibly large)
magnetic field, $\hat H(\lambda) = \lambda \hat J_z$, and consider the microcanonical ensemble of states with expectation $E=0$. 
Besides the state with null angular momentum (the only state contributing to the standard microcanonical ensemble), superposition
containing both the up and down states now contribute to the ensemble as well.} For this reason various authors claim  that
the ensemble in Eq. (\ref{eq:microcan}) provides a more realistic description of the thermodynamics of isolated systems 
\cite{Reimann08PRL101,Reimann07PRL99,Fine09PRE80,Ji11PRL107}.
Another pleasing property of this ensemble is that, at variance with the standard microcanonical ensemble,
it does not require a dense energy spectrum, and can therefore be well applied to small quantum systems with well
separated energy levels. Indeed, the ensemble depends continuously on the real parameters $E,\lambda$, which
makes the derivation of the associated thermodynamics straightforward also in case of small systems
\cite{Brody07PRSA463}. Ref. \cite{Ji11PRL107} shows that this ensemble well describes the statistics of 
a small thermally isolated system after repeated non-adiabatic perturbations.

\subsection{Canonical wave function ensemble}
In the canonical wave function ensemble \cite{Brody98JMP98,Jona-Lasinio06AIPCP844},
wave functions are weighted with the Gibbs factor 
$e^{-\beta h(\mathbf x, \mathbf p; \lambda)}$:
\begin{eqnarray}
\rho_{c}(\mathbf x, \mathbf p; \beta,\lambda)= \frac{e^{-\beta h(\mathbf x, \mathbf p; \lambda)}
 \delta(1- |\mathbf x + i\mathbf p|^2 )}{Z(\beta,\lambda)}
  \label{eq:can}
\end{eqnarray}
where 
\begin{eqnarray}
Z(\beta,\lambda) = \int d\mathbf x d\mathbf p\, e^{-\beta h(\mathbf x, \mathbf p; \lambda)} \delta(1- |\mathbf x + i\mathbf p|^2 )
 \label{eq:Z}
\end{eqnarray}
Note again the formal similarity with the classical canonical ensemble. 
In Ref. \cite{Jona-Lasinio06AIPCP844} this ensemble is called the Schr\"odinger-Gibbs 
ensemble.
According to \cite{JonaLasinio00InBook} this ensemble can give realistic predictions 
in case of mesoscopic systems where robust coherence phenomena are involved.

\section{Rationale for wave function ensembles}
A criterion for establishing the goodness of a statistical ensembles as a candidate model of
equilibrium thermodynamics is whether the ensemble is invariant under the time evolution.
As will become clearer in the next section this is indeed the case for the canonical and microcanonical
wave function ensemble.

Another criterion, which traces back to Boltzmann \cite{GallavottiBook},
is whether the ensemble endows the parameter space with a ``thermodynamic structure''.
To be more explicit, given a statistical ensemble $\rho(\Gamma, X_i)$,
(defined on a phase space $\Gamma$ and on a parameter space $X_i$), 
one checks whether there exist an integrating factor $\gamma(X_i)$, such that
\begin{equation}
\gamma \, \delta Q = \quad $exact differential$
\label{eq:heatTheo}
\end{equation}
where $\delta Q$ is the heat differential as calculated in the ensemble.
This equation is known as the {\em heat theorem}, and is the most fundamental equation
of thermodynamics. Prominent examples of textbooks that take this viewpoint in establishing 
the foundations of quantum statistics are those of  Schr\"odingier \cite{SchrodingerBook}, and
Khinchin \cite{Khinchin51Book}.\footnote{Interestingly both books also advocate the use of wave function ensembles.}

To calculate $\delta Q$ use the standard formula
\begin{equation}
\delta Q = d E +  F d\lambda
\label{eq:deltaQ}
\end{equation}
where
\begin{eqnarray}
E = \langle \hat H \rangle\\
F = -   \left \langle \frac{\partial \hat H}{\partial \lambda} \right\rangle
\end{eqnarray}
denote the ensemble averages of energy and of  the generalized force conjugated to the external parameter.
Note that in case of a single parameter $\lambda$, mathematics ensures that an integrating factor 
always exists. A differential form in two dimensions (i.e., $E$ and $\lambda$, in Eq. (\ref{eq:deltaQ})),
always admits an integrating factor. 
However, the system Hamiltonian may depend on many external parameters $\lambda_i$, hence
 $\delta Q = dE + \sum_i F_i d\lambda_i$,
which makes the question of the existence of an integrating factor non-trivial.

\subsection{Canonical case}
In the canonical case we have 
\begin{eqnarray}
E= E(\beta, \lambda) =  \int d\mathbf x d\mathbf p\,  \rho_c(\mathbf x, \mathbf p; \beta, \lambda) h(\mathbf x, \mathbf p;\lambda)\\
F = F(\beta, \lambda) =  -\int d\mathbf x d\mathbf p\,  \rho_c(\mathbf x, \mathbf p;\beta, \lambda) \frac{\partial h(\mathbf x, \mathbf p;\lambda)}{\partial \lambda}
\end{eqnarray}
In this case $\beta$ is an integrating factor for $\delta Q$ and $S_c(\beta,\lambda)= \beta E(\beta,\lambda)+\ln Z(\beta,\lambda)$
is the associated generating function. The argument follows step by step the classical derivation \cite{Campisi07PHYSA385}, which
can be repeated without modifications. The partial derivatives of $S_c(\beta,\lambda)$ are:
\begin{eqnarray}
\frac{\partial S_c}{\partial \beta} = E + \beta \frac{\partial E}{\partial \beta} + \frac{1}{Z}\frac{\partial Z}{\partial \beta}= \beta \frac{\partial E}{\partial \beta}\\
\frac{\partial S_c}{\partial \lambda} =\beta \frac{\partial E}{\partial \lambda} + \frac{1}{Z}\frac{\partial Z}{\partial \lambda}= 
\beta \frac{\partial E}{\partial \lambda} + \beta F
\end{eqnarray}
therefore 
\begin{eqnarray}
dS_c = \beta \left( \frac{\partial E}{\partial \beta}d \beta + \frac{\partial E}{\partial \lambda} d\lambda + Fd\lambda \right)
= \beta (dE + Fd\lambda) = \beta \delta Q
\end{eqnarray}
The derivation can be straightforwardly repeated in the case of many parameters.
We remark that there are however infinitely many integrating factors for $\delta Q$. So having found one does not ensure by itself
that it can be interpreted as inverse temperature, and that the associated generator of the exact differential can be interpreted
as entropy. Take for example $g(\beta,\lambda)=f(S_c(\beta,\lambda))$ with any monotonic function $f$. Then $d g = f'(S_c(\beta,\lambda))d S_C = f'(S_c(\beta,\lambda))\beta \delta Q$, where $f'$ is the derivative of $f$. This says that $f'(S_c(\beta,\lambda))\beta$, is also an integrating factor for $\delta Q$.
In order to pick the ``thermodynamic'' integrating factor, we need an extra ingredient.
We thus further require that the entropy be additive. Namely, if two non interacting and non-entangled systems have separately the entropies $S_1$ and $S_2$, the entropy of the total system should be $S_1+S_2$. The requirement of non-entanglement is very crucial here. It restricts the Hilbert space of the compound system, from a tensor product of dimension $N_1 N_2$ to the direct product of dimension $N_1+N_2$. In this ``classical'' phase space the canonical wave function distribution of the compound system reduces to the product of the canonical wave function distributions for each subsystem, so does the partition function $Z$. Noting that the energy is additive, it follows that $S_c(\beta,\lambda)$ is additive as well, which singles it out as a good candidate for thermodynamic entropy. Accordingly $\beta$ is the inverse temperature.

\subsection{Microcanonical case}
In the microcanonical case
\begin{eqnarray}
E =  \int d\mathbf x d\mathbf p\,  \rho_\mu(\mathbf x, \mathbf p;E, \lambda) h(\mathbf x, \mathbf p;\lambda)\\
F = F(E, \lambda) =  -\int d\mathbf x d\mathbf p\,  \rho_\mu(\mathbf x, \mathbf p;E,\lambda) \frac{\partial h(\mathbf x, \mathbf p;\lambda)}{\partial \lambda}
\end{eqnarray}
An integrating factor for $\delta Q$ is in this case the function $\Omega(E,\lambda)/\Phi(E,\lambda)$, where, in analogy with classical mechanics
\begin{eqnarray}
\Phi(E, \lambda) = \int d\mathbf x d\mathbf p\,  \theta(E-h(\mathbf x, \mathbf p; \lambda))
 \delta(1- |\mathbf x + i\mathbf p|^2 )
\end{eqnarray}
denotes the volume of physical Hilbert space with energy expectation below $E$. As in classical mechanics, we have $\Phi(E,\lambda)=\int_{E_0}^E \Omega(E',\lambda)dE'$, where $E_0$ is the ground state energy. The symbol $\theta$ denotes the Heaviside step function. 
The proof follows, \emph{mutatis mutandis}, the classical argument (the generalized Helmholtz theorem) \cite{Campisi05SHPMP36}, which can be repeated also with many external parameters.
The generating function associated with the integrating factor $\Omega/\Phi$ is $S_\mu(E,\lambda)= \ln \Phi(E, \lambda)$. 
In this case the requirement of additivity does not seem to single $S_\mu(E,\lambda)$ so straightforwardly as in the canonical case.
The reason is that, unlike the exponential, the theta function does not factorize in the product of two theta functions.
Classically this problem can be easily circumvented upon noticing that the integrating factor $\Omega/\Phi$ equals the average kinetic energy per degree of freedom (equipartition theorem \cite{Khinchin49Book}), which singles it out as the thermodynamic temperature. In quantum mechanics however there is no equipartition theorem to help us. We leave the resolution of this question to future studies.

It should be remarked that our present analysis contrasts with Ref. \cite{Brody07PRSA463}, where thermodynamics was derived from the logarithm of the density of states, namely  $\ln \Omega(E,\lambda)$. We remark that this choice does not comply with the heat theorem, Eq. (\ref{eq:HamEqs}), namely, there does not exist, in general  a function $\gamma(E,\lambda)$, such that $\gamma(E,\lambda) \delta Q$ would equal the differential of $\, \ln \Omega(E,\lambda)$. This very same question appears also at the classical level, where it has been long ignored due to the fact that in most cases of interest the ``surface entropy'' (logarithm of the density of states) and the ``volume entropy'' (logarithm of the integrated density of states), give practically undistinguishable results for sufficiently large systems \cite{Campisi05SHPMP36,Dunkel13arXiv}.

\section{Fluctuation relations}
Fluctuation relations for the wave function ensembles follow straightforwardly
upon noticing that in the $(\mathbf x, \mathbf p)$ representation, the Schr\"odinger equation
\begin{equation}
i \hbar \dot{\mathbf c} = \mathbf H(\lambda_t) \mathbf c
\end{equation}
assumes the form of classical Hamilton's equation
\begin{eqnarray}
\dot{\mathbf x} = \frac{\partial }{\partial \mathbf p} h(\mathbf x, \mathbf p; \lambda_t)\\
\dot{\mathbf p} = -\frac{\partial }{\partial \mathbf x} h(\mathbf x, \mathbf p; \lambda_t)
\label{eq:HamEqs}
\end{eqnarray}
with the function $h(\mathbf x, \mathbf p;\lambda_t)$ being the generator of the dynamics \cite{Strocchi66RMP38}.
In analogy with the classical case, we introduce the following notion of quantum work
\begin{eqnarray}
w = h(\mathbf x_\tau, \mathbf p_\tau; \lambda_\tau) - h(\mathbf x, \mathbf p; \lambda_0)
\label{eq:work}
\end{eqnarray}
where $(\mathbf x_\tau, \mathbf p_\tau)$ denotes the evolved of $(\mathbf x, \mathbf p)$,
according to Hamilton's equations (\ref{eq:HamEqs}). Physically, $w$ is the change
in the expectation of the Hamilton operator $\hat H$, due to the evolution of the wave function $|\psi\rangle$,
 see Eq. (\ref{eq:w-nomeas}). Note that $w$ can be expressed as an integrated power:
\begin{eqnarray}
w = \int_0^\tau dt \dot \lambda_t \frac{\partial h(\mathbf x_t, \mathbf p_t, \lambda_t)}{\partial \lambda_t}
\label{eq:work-int}
\end{eqnarray}

In equilibrium, namely for a constant $\lambda$, energy conservation and 
Liouville theorem ensure that surfaces
of constant energy expectation in the physical Hilbert space  will be mapped onto themselves by the time evolution, implying that,
as anticipated, the canonical and microcanonical wave function ensembles are stationary \cite{JonaLasinio00InBook}.

The probability that the work $w$ be performed on a system
prepared in a wave function ensemble $\rho(\mathbf x, \mathbf p)$ can be written as
\begin{eqnarray}
p(w) = \int d \mathbf x d \mathbf p\,  \rho(\mathbf x, \mathbf p)
\delta(w-h(\mathbf x_\tau, \mathbf p_\tau, \lambda_\tau) + h(\mathbf x, \mathbf p, \lambda_0)  )
 \end{eqnarray}
Noticing that the evolution (\ref{eq:HamEqs}) conserves the normalization, 
${|\mathbf x_\tau +i\mathbf p_\tau|^2=|\mathbf x +i\mathbf p|^2=1}$ (unitarity of quantum evolution) and
is volume preserving, $d \mathbf x_\tau d \mathbf p_\tau = d \mathbf x d \mathbf p $ (classical Liouville theorem), one can repeat step by step the
derivations of classical microcanonical \cite{Cleuren06PRL96} and canonical \cite{Jarzynski11ARCMP2}
fluctuation relations, upon requiring that the Hamilton operator is time reversal invariant.\footnote{Formally
that means that at each time $t$, the Hamilton operator $\hat H(\lambda_t)$ commutes with time-reversal operator $\Theta$, which
changes the sign of momenta and leaves spatial coordinates unchanged \cite{Messiah62Book}}

In the microcanonical case one obtains:
\begin{eqnarray}
\frac{p_E(w)}{\widetilde p_{E+w}(-w)} = \frac{\Omega(E+w,\lambda_\tau)} {\Omega(E,\lambda_0)}
\end{eqnarray}
where $p_E(w)$ is the probability of doing work $w$ when the initial state is randomly drawn
from the distribution  $\rho_\mu(\mathbf x, \mathbf p; E, \lambda_0)$  under the driving protocol $\lambda_t$
$t \in [0,\tau]$ , and 
$\widetilde p_{E+w}(-w)$ is the probability of doing work $-w$ when the initial state is randomly drawn
from ${\rho_\mu(\mathbf x, \mathbf p; E+w, \lambda_\tau)}$ under the protocol $\lambda_{\tau-t}$,
$t \in [0,\tau]$.

In the canonical case one obtains:
\begin{eqnarray}
\frac{p(w)}{\widetilde p(-w)} = \frac{Z(\beta,\lambda_\tau)} {Z(\beta,\lambda_0)} e^{\beta w}= e^{-\beta(\Delta F - w)}
\end{eqnarray}
where $p(w)$ is the probability of doing work $w$ when the initial state is randomly drawn
from the distribution  $\rho_c(\mathbf x, \mathbf p; \beta, \lambda_0)$  under the driving protocol $\lambda_t$
$t \in [0,\tau]$ , and 
$\widetilde p(-w)$ is the probability of doing work $-w$ when the initial state is randomly drawn
from ${\rho_c(\mathbf x, \mathbf p; \beta, \lambda_\tau)}$ under the protocol $\lambda_{\tau-t}$,
$t \in [0,\tau]$.
In analogy with the classical case we have introduced the notation 
$\Delta F = F(\beta, \lambda_\tau)-F(\beta, \lambda_0)$, with $F(\beta, \lambda) = -\beta^{-1}\ln Z(\beta, \lambda)$.
We stress that this free energy $F(\beta,\lambda)$ may considerably differ from the
usual free energy  $F_{st}(\beta, \lambda)=-\beta^{-1}\ln\Tr e^{-\beta \hat H(\lambda)}$, see Fig. \ref{fig:Fig1}.a.

\section{Illustrative example}
To better clarify the differences and similarities between the standard quantum fluctuation relations and the
quantum fluctuation relations for wave function ensembles, we consider the Landau-Zener(-St\"{u}ckelberg-Majorana)
\cite{Landau32PZS2,Zener32PRSA137,Stueckelberg32HPA5,
Majorana32NC9} problem
\begin{equation}
 \hat H(\lambda_t)= \lambda_t\sigma_z +\Delta \sigma_x\, ,\qquad \lambda_t = v t/2 \, .
\label{eq:H-LZ}
\end{equation}
It governs the dynamics of a two-level quantum system whose 
energy separation, $vt$, varies linearly in time, and whose states 
are coupled via the interaction energy $\Delta$. For example, a
spin-$1/2$ particle with magnetic moment $\mu$ in a magnetic field
$\vec{B}_t = - (\Delta/\mu, 0,  v t/2\mu)$.
Here, $\sigma_x$ and $\sigma_z$ denote Pauli matrices.

\begin{figure}[t]
		\includegraphics[width=\textwidth]{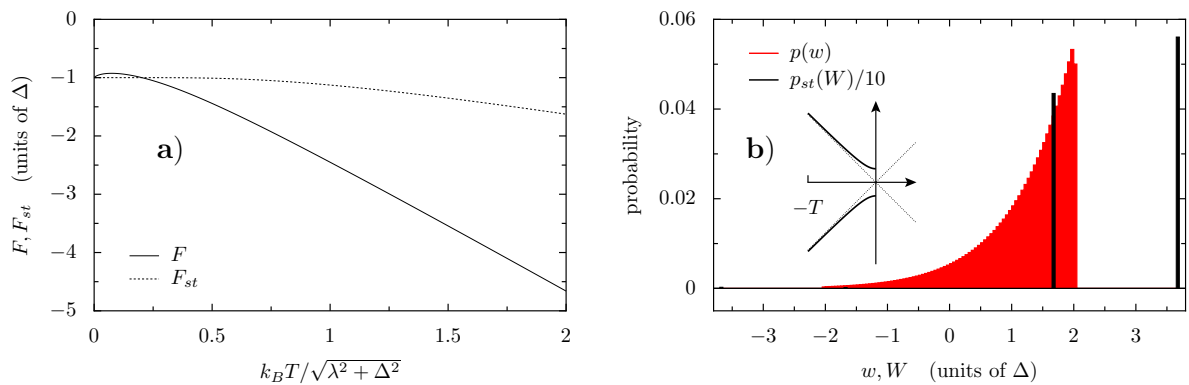}
		\caption{Panel a: Free energy of a two-level-system described by the Hamiltonian (\ref{eq:H-LZ})
		with a fixed $\lambda$, as computed in the canonical (Schr\"odinger-Gibbs) wave function ensemble, Eq. (\ref{eq:can}), and in the standard
		Gibbs canonical ensemble. Panel b: Probability histograms of work $w$, Eq. (\ref{eq:work}), and standard work $W$, Eq.	(\ref{eq:w-two-mess}), as computed in the canonical  (Schr\"odinger-Gibbs) wave function ensemble, Eq. (\ref{eq:can}), and in the standard Gibbs canonical ensemble, respectively. The standard work probability $p_{st}(W)$ is rescaled by a factor 10 for better visualization. Inset: sketch of the driving protocol, i.e. a half Landau-Zener sweep. The parameters used are: $\beta=\Delta^{-1},v=\Delta^2/\hbar, T= 5 \hbar/\Delta$.
}
		\label{fig:Fig1}
\end{figure}

Let us assume the two-level system is in a state described by the
canonical wave function ensemble, Eq. (\ref{eq:can}). Let $\mathbf{c}=(a,b)^T$, with
$a,b \in \mathbb{C}$, denote a point in the Hilbert space (a wave function). 
The energy expectation $h(a,b,\lambda)$ over the state $\mathbf{c}$ reads
$
h(a,b;\lambda) = \lambda(|a|^2- |b|^2) +  \Delta (a^* b+ a b^*)
$,
where $^*$ denotes complex conjugation.
Accordingly, the  partition function reads:
\begin{eqnarray}
Z(\beta,\lambda) = \int da\, db\,  e^{-\beta [\lambda(|a|^2- |b|^2) +  \Delta (a^* b+ a b^*)
]} \delta(1- |a|^2- |b|^2 )
 \label{eq:Z-LZ}
\end{eqnarray}
As is well known, the projective Hilbert space of a two-level system can be mapped onto a 
sphere of unit radius in $ \mathbb{R}^3$, the Bloch sphere. Accordingly, the 
partition function $Z(\beta,\lambda)$ can be expressed as an integral over the Bloch sphere. This is accomplished 
by the following change of variables, $a=e^{i\phi}r \cos \gamma/2, b= e^{i\phi} r e^{i\delta}\sin\gamma/2$,
where $r\in[0,\infty), \phi \in [0,2\pi], \gamma \in [0,\pi],  \delta \in [0,2\pi]$, leading to:
\begin{eqnarray}
Z(\beta,\lambda) &=& \frac{1}{8} \int dr^2\, d\phi \, d\delta\, d\gamma\, \sin \gamma \,r^2 e^{-\beta r^2 [\lambda \cos\gamma +\Delta \sin\delta \, \sin \gamma]} \delta(1- r^2 )\nonumber \\
 &=& \frac{\pi}{4} \int d\delta\, d\gamma\, \sin \gamma \, e^{-\beta [\lambda \cos\gamma +\Delta \sin\delta \, \sin \gamma]} 
 \label{eq:Z-LZb}
\end{eqnarray}
where $\gamma,\delta$ are the Bloch angles. To perform the integration we first consider the case
$\Delta=0$. Physically this corresponds to a spin-$1/2$ particle in a 
magnetic field pointing in the negative $z$ direction with intensity $\lambda/\mu$. 
By the change of variable $y=\cos \gamma$ we obtain, for $\Delta=0$, $Z= \pi^2 \sinh(\beta\lambda)/(\beta\lambda)$.
When $\Delta \neq 0$, this corresponds to a magnetic field oriented along some direction $\hat\mathbf{n}$ and an intensity
$\sqrt{\lambda^2+\Delta^2}/\mu$. Because of spatial isotropy, the partition function can only depend on the intensity of the field
and not on its orientation, hence we obtain,
\begin{eqnarray}
Z(\beta,\lambda) =  \pi^2 \frac{\sinh(\beta \sqrt{\lambda^2+\Delta^2})}{\beta \sqrt{\lambda^2+\Delta^2}}\, .
\end{eqnarray}
This expression should be contrasted with the standard expression 
$Z_{st}(\beta,\lambda) = \Tr\,  e^{-\beta \hat H(\lambda)}= 2\cosh(\beta \sqrt{\lambda^2+\Delta^2})$.
Figure \ref{fig:Fig1}.a shows a comparison of the resulting free energies, $F=-\beta^{-1}\ln Z$, $F_{st}=-\beta^{-1}\ln Z_{st}$.
As already highlighted in Ref. \cite{Brody98JMP98}
they give rise to distinct thermodynamics.

It is worth stressing that, just like the standard ensemble, the wave function ensemble is a
mixed state which can, accordingly, be represented by a density matrix \cite{Jona-Lasinio06AIPCP844}:
$\hat \rho(\beta, \lambda)= \int d\mathbf{x}d\mathbf{p} \rho(\mathbf{x},\mathbf{p},\beta,\lambda) (\mathbf{x}-i\mathbf{p})(\mathbf{x}+i\mathbf{p})^T$. In the present case it reads, in the $\sigma_z$ basis
\begin{equation}
\hspace{-2cm}\hat \rho(\beta,\lambda) = \frac{\pi}{4} \int d\delta\, d\gamma\, \sin \gamma \, \frac{e^{-\beta [\lambda \cos\gamma +\Delta \sin\delta \, \sin \gamma]}}{Z(\beta,\lambda)} \left(\begin{array}{cc} \cos^2(\gamma/2) & \sin \gamma e^{-i \delta}/2\\  \sin \gamma e^{i \delta}/2 & \sin^2(\gamma/2)\end{array}\right)
\end{equation}
In the case $\Delta=0$ we get \cite{Brody98JMP98}
\begin{equation}
\hspace{-2cm}\hat \rho(\beta,\lambda) =\frac{1}{2} \left(\begin{array}{cc} 1 + 1/(\beta \lambda) - \coth(\beta \lambda)& 0\\  0 & 1 - 1/(\beta \lambda) + \coth(\beta \lambda))\end{array}\right),\quad \Delta=0\, .
\end{equation}
This density matrix should be contrasted with the standard canonical density matrix $\hat \rho_{st}(\beta,\lambda) = \mbox{diag}(e^{-\beta \lambda}, e^{\beta \lambda})/Z_{st}(\beta, \lambda)$.
By replacing $\lambda$ with $\sqrt{\lambda^2+\Delta^2}$, one gets the density matrix for the case $\Delta \neq 0$, in the corresponding energy eigenbasis.

In Fig. \ref{fig:Fig1}.b we report results concerning the work statistics. We considered here a ``half'' Landau-Zener sweep, i.e., Eq. (\ref{eq:H-LZ}) from time $t=-T$, to time $t=0$, see the inset of Fig. \ref{fig:Fig1}.b.
The unitary quantum evolution operator can be expressed in terms of special functions \cite{Vitanov99PRA59,Campisi11PRE83}.
The figure shows both the statistics $p(w)$ originating from the expression of work in Eq. (\ref{eq:work}) in the 
canonical wave function ensemble, Eq. (\ref{eq:can}), and the standard work statistics $p_{st}(W)$ originating from the 
two-measurement expression of work in Eq. (\ref{eq:w-two-mess}) in the
standard canonical ensemble $e^{-\beta \hat H(\lambda)}/Z_{st}(\beta,\lambda)$.
\footnote{To be more precise, Fig. \ref{fig:Fig1}.b shows the quantities $\int _{w-d/2}^{w+d/2}p(w')dw'$, and $\int _{W-d/2}^{W+d/2}p_{st}(W')dW'$, (with $d$ the width of the bars), i.e., discrete versions of $p(w)$ and $p_{st}(W)$. $p_{st}(W)$ is rescaled by a factor 10 in  Fig. \ref{fig:Fig1}.b, for a better visualization.}
Note the prominent difference that the wave function work pdf $p(w)$ is a smooth function whereas the standard 
work pdf $p_{st}(W)$ is a discrete sum of 4 Dirac deltas  \cite{Campisi11RMP83} (the two most left peaks of $p_{st}(W)$ are barely visible in Fig. \ref{fig:Fig1}.b). Note also that the support of $p(w)$ is smaller than the support of $p_{st}(W)$. Stronger driving (i.e. larger $v$'s) result in broader
distributions $p(w)$. The support of $p(w)$ cannot however become wider that that of $p_{st}(W)$, which, independent of $v$, is given by
$[-\sqrt{(vT/2)^2+\Delta^2}-\Delta,\sqrt{(vT/2)^2+\Delta^2}+\Delta]$. 

Notwithstanding their differences both distributions satisfy formally equivalent fluctuation relations.
To better clarify this, let us focus on the average exponentiated work. As predicted by the theory and confirmed
by our numerical calculation, we have:
\begin{eqnarray}
\langle e^{-\beta w}\rangle = \int dw p(w) e^{-\beta w} = e^{-\beta \Delta F} \\
\langle e^{-\beta W}\rangle_{st} = \int dW p_{st}(W) e^{-\beta W} = e^{-\beta \Delta F_{st}} 
\end{eqnarray}
That is, both work pdf's satisfy the Jarzynski equality, each with the free energy calculated in the respective ensemble.
Likewise for the Tasaki-Crooks fluctuation theorem.

\section{Concluding remarks}
We have obtained fluctuation relations for microcanonical and canonical wave function 
ensembles. They look exactly as the standard relations, but substantially differ from them because
they involve a notion of work as the change in the
expectation of the energy, rather than the difference of two eigenvalues emerging from 
quantum collapses. These ensembles in fact have been proposed in a framework where one is interested in the 
the expectation of quantum observables \cite{Jona-Lasinio06AIPCP844}.
As highlighted with the illustrative example, this notion of work gives rise
to smooth work probability densities, in stark contrast with the discrete standard probability densities.
Also it gives information about the equilibrium ``free energy'' (``entropy'') as calculated in the 
canonical (microcanonical) wave function ensemble. These substantially differ from their standard 
counterpart, see Fig. \ref{fig:Fig1}.

Other authors are currently developing alternative formulations of
quantum fluctuation relations which do not rely on quantum collapses. Among them is the work of 
Ref. \cite{Deffner13EPL103} which presents a study of entropy production based on the Wigner 
representation of quantum states.

We have expressed some considerations regarding the rational foundations of the 
wave function ensembles. Further investigation is certainly necessary in order to
reach a more satisfactory understanding of the physical basis for these ensembles. 
One question to be pursued regards the 
lack of ergodicity of the Hamiltonian flow on the surface of constant energy 
expectation in the physical Hilbert space, which marks a stark distinction with the classical case.
Another important question that deserves further study is whether these ensembles
converge to the usual statistical ensembles in some limit, e.g. classical, and/or thermodynamic limit.
Experiments will have the final word in regard to their  scope of applicability. 
Certainly they have proved very important in recent advancements
in the foundations of quantum statistics \cite{Goldstein06PRL96,Popescu06NATPHYS2}.

\section*{Ackowledgements}
This work was supported by the German Excellence Initiative ``Nanosystems Initiative Munich (NIM)''.

\section*{References}
\bibliographystyle{iopart-num}

\providecommand{\newblock}{}

\end{document}